 \documentclass[10pt,pre,twocolumn,preprintnumbers,amsmath,amssymb]{revtex4-1}
 
\usepackage{stmaryrd} 
\usepackage{hyperref} 

 \usepackage{bm}
 \usepackage{graphicx}
 \usepackage{latexsym}

\newcommand{\grad}{ \mbox{\boldmath\(\nabla\)}}
\newcommand{\divv}{ \mbox{\boldmath\(\nabla\cdot\)}}
\newcommand{\curl}{ \mbox{\boldmath\(\nabla\times\)}}

\newcommand{\dotv}{  \mbox{\boldmath\(\cdot\)} }

\newcommand{\cross}{  \mbox{\boldmath\(\times\)} }
\newcommand{\sgn} {\mathrm{sgn}\,}
\renewcommand{\Re} {\mathrm{Re}\,}
\renewcommand{\Im} {\mathrm{Im}\,}
\newcommand{\half}{\mbox{\small\(\frac{1}{2}\)} }

\newcommand{\Eq}[1]{Eq.~(\ref{#1})}

\newcommand{\vrm}[1]{\bm{\mathrm{#1}}} 
\newcommand{\vsf}[1]{\bm{\mathsf{#1}}} 

\newcommand{\jump}[1]{\left\llbracket  #1 \right\rrbracket} 

\begin{document}

 \title{
Plasmoid solutions of the Hahm--Kulsrud--Taylor equilibrium model
 }

\author{R.~L. Dewar}
 \email{robert.dewar@anu.edu.au}
 \affiliation{Plasma Research Laboratory, Research School of Physics \& Engineering, The Australian National University, Canberra, ACT 0200, Australia}
 \affiliation{Graduate School of Frontier Sciences, University of Tokyo, Kashiwa, Chiba 277-8561, Japan}

\author{A. Bhattacharjee}
 \email{abhattac@pppl.gov}
 \affiliation{Princeton Plasma Physics Laboratory, PO Box 451, Princeton NJ 08543, USA}

\author{R.~M. Kulsrud}
 \email{rkulsrud@pppl.gov}
 \affiliation{Princeton Plasma Physics Laboratory, PO Box 451, Princeton NJ 08543, USA}

\author{A.~M. Wright}
 \affiliation{The Australian National University, Canberra, ACT 0200, Australia}

\date{\today}

\begin{abstract}
The Hahm--Kulsrud (HK) [T.~S. Hahm and R.~M. Kulsrud, Phys. Fluids {\bf 28}, 2412 (1985)] solutions for a magnetically sheared plasma slab driven by a resonant periodic boundary perturbation illustrate fully shielded (current sheet) and fully reconnected (magnetic island) responses. On the global scale, reconnection involves solving a magnetohydrodynamic (MHD) equilibrium problem. In systems with a continuous symmetry such MHD equilibria are typically found by solving the Grad--Shafranov equation, and in slab geometry the elliptic operator in this equation is the 2-D Laplacian. Thus, assuming appropriate pressure and poloidal current profiles, a conformal mapping method can be used to transform one solution into another with different boundary conditions, giving a continuous sequence of solutions in the form of partially reconnected magnetic islands (plasmoids) separated by  Syrovatsky current sheets. The two HK solutions appear as special cases.
\end{abstract}

\maketitle

\section{Introduction}
\label{sec:Intro}

Recently, there has been renewed interest in the secondary tearing instability of high-Lundquist-number current sheets \cite{Loureiro_Schekochihin_Cowley_07}, called the ``plasmoid instability'' \cite{Bhattacharjee_Huang_Yang_Rogers_09}. Numerical simulations, supported by heuristic scaling arguments \cite{Huang_Bhattacharjee_10}, demonstrate that if the Lundquist number ($S$) based on the length of the current sheet exceeds a threshold \cite{Biskamp_86,Parker_Dewar_Johnson_90}, the instability breaks up a Sweet--Parker current layer into a sequence of magnetic islands separated by segments of current sheets, and evolves into a new nonlinear regime of reconnection in which the reconnection rate becomes nearly independent of $S$ \cite{Huang_Bhattacharjee_10,Lapenta_08,Daughton_etal_09,Cassak_Shay_Drake_09,Loureiro_etal_09}. These simulation results suggest that there might exist partially reconnected plasmoid solutions of the magnetostatic equilibrium equations in which plasmoids exist, separated by segments of current sheets. In this paper, we show that such solutions can indeed be constructed within the framework of the Hahm--Kulsrud--Taylor (HKT) model, described below.

The HKT model, developed by Hahm and Kulsrud \cite{Hahm_Kulsrud_85} following a suggestion by J.~B. Taylor, considers the response of a plasma slab with a sheared unperturbed magnetic field to a resonant perturbation applied at the boundaries $x = \pm a$. In Cartesian coordinates $x,y,z$ the magnetic field $\vrm{B}$ is represented as $\grad z \cross\grad \psi(x,y) + B_z(x,y) \grad z$. The unperturbed ``poloidal'' flux function $\psi$ is $\psi_0(x) = B_a x^2/2a$, where the constant $B_a$  is the strength of the poloidal magnetic field at the boundary $x = a$.

In Ref.~\onlinecite{Hahm_Kulsrud_85} the ``toroidal'' field $B_z$ was assumed to be effectively constant and much larger than $B_a$ to allow incompressibility to be assumed during a discussion of reconnection dynamics, but in this paper we will be concerned only with finding static equilibrium solutions so this assumption is not necessary. Instead,  we regard the $B_z$ profile function $F(\psi)$ in the Grad--Shafranov equation as free to choose,  and assume it is chosen so that $B_z^2/2\mu_0 + p$ is \emph{linear} in $\psi$, where $p(\psi)$ is the pressure (and, for SI units, $\mu_0$ is the permeability of free space). This makes the Grad--Shafranov equation linear (though inhomogeneous) allowing analytic solutions to be obtained.

\begin{figure}[htbp]
   \centering
		\includegraphics[width = 0.45\textwidth]{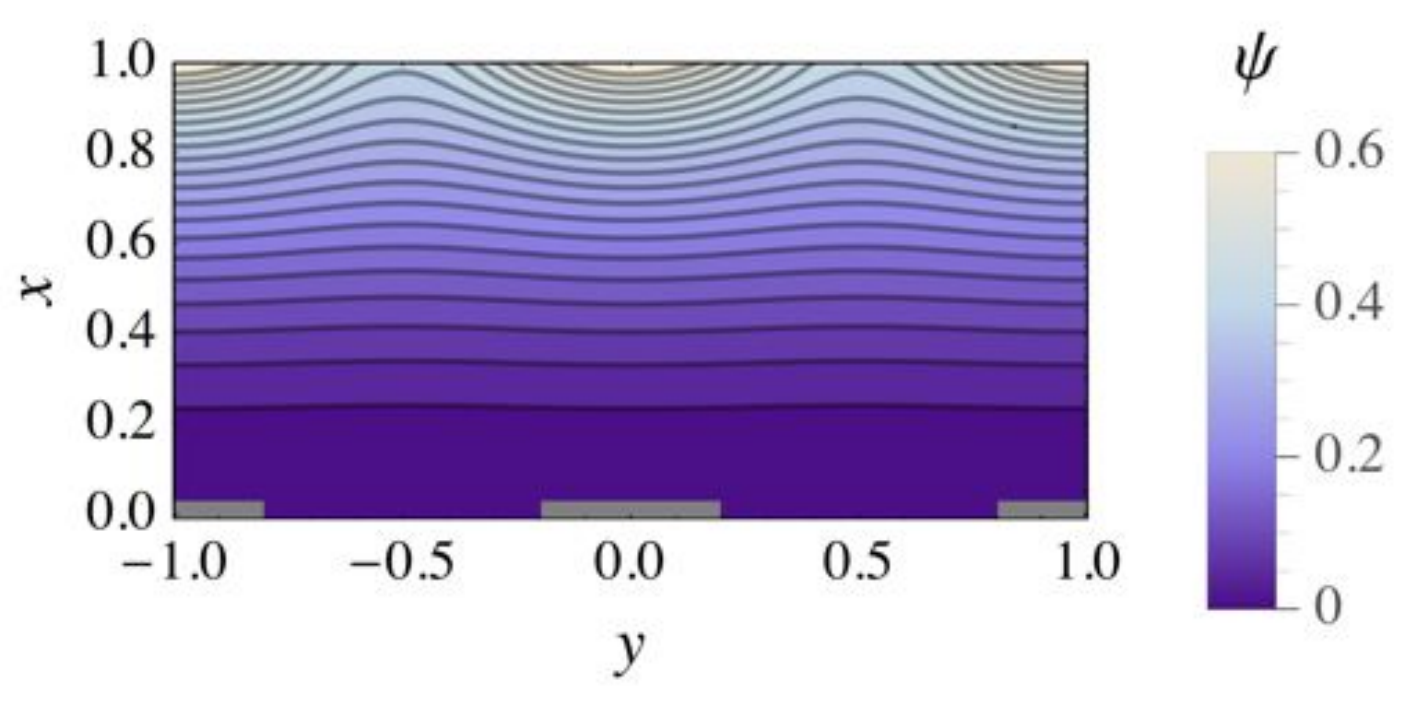} 
\caption{Contours of $\psi = \psi_0+\alpha\tilde\psi$ for the standard illustrative values used throughout this paper: $\alpha=0.1$, units such that $a=B_a=1$, and wavelength $\lambda$ chosen to equal $a$, so $k=2\pi$. This illustrates the effective rippling of the upper boundary (axes scales equal). Cuts corresponding to the case $L = \lambda/6$ are also shown as thick gray lines on the $y$-axis. (Color online.)}
\label{fig:Boundary}
\end{figure}

The perturbed flux function at the boundaries is assumed to be
\begin{equation}\label{eq:bndcond}
\begin{split}
	\psi(\pm a,y)  & = \half a B_a + B_a\delta\cos ky  \\
	  			& \equiv a B_a\left(\half + \alpha\cos ky\right) \;,
\end{split}
\end{equation}
where  $\alpha \equiv \delta/a$ is a dimensionless parameter measuring the strength of the perturbation,
$\delta$ being the amplitude of a notional small boundary ripple of wavelength $\lambda = 2\pi/k$, the given planar boundary conditions being, to linear order in $\alpha \ll 1$, equivalent to a symmetric geometric rippling of perfectly conducting bounding walls.

This rippling is illustrated in Fig.~\ref{fig:Boundary}), though the value of $\alpha = 0.1$ we use in our standard illustrative case is rather too large for the ripples to be even approximately sinusoidal. However, in this paper we are do not really need to ripple the boundary and take the HKT boundary conditions \Eq{eq:bndcond} as \emph{exact}, so that $\alpha$ need not be infinitesimal. Thus, because of the assumed linearity of the Grad--Shafranov equation mentioned above, we have exact linearity and can write
\begin{equation}
	\psi = \psi_0(x) + \alpha\tilde\psi(x,y) \;,
	\label{eq:psitildedef}
\end{equation}
where $\tilde\psi$ is independent of $\alpha$ and obeys the boundary conditions $\tilde\psi(\pm a,y) = a B_a\cos ky$.

Ideal magnetohydrodynamics (MHD) cannot predict the timescale for magnetic reconnection (magnetic field changes that violate the topological frozen-in-flux condition). However, on a long enough length scale and a short enough time scale, the intermediate states in the evolution of driven \cite{Uzdensky_Kulsrud_Yamada_96,Uzdensky_Kulsrud_97,Uzdensky_Kulsrud_00,Yamada_Kulsrud_Ji_10} or spontaneous \cite{Wang_Bhattacharjee_95} magnetic reconnection can be described as a continuous sequence of ``global''  MHD equilibrium states (i.e. states that satisfy the boundary conditions and internal force balance). In the present paper we do not seek to describe the reconnection process in detail, merely to find analytically solvable Grad--Shafranov equilibria that plausibly illustrate a possible reconnection scenario. 

In Sec.~\ref{sec:GS} it is pointed out that the Grad--Shafranov equation in slab geometry can include current sheets in two ways---either as a superposition of $\delta$-function current-density sources or as cuts in the $x,y$ plane, the latter being the viewpoint used in this paper. Force balance provides the boundary conditions on the cuts. Current profiles are given such that the Grad--Shafranov equation in slab geometry becomes a linear Poisson equation and our definition of the HKT equilibrium problem is made precise.

Hahm and Kulsrud  \cite{Hahm_Kulsrud_85} found two \emph{exact} MHD equilibrium solutions, one involving a full current sheet covering the plane $x = 0$ and one describing a magnetic island with no current sheet. The current sheet solution may be viewed as representing how a shielding current \cite[e.g]{Boozer_Pomphrey_11} initially arises in order to prevent reconnection after a resonant perturbation is turned on,  the magnetic island solution being interpreted as the end state after a sufficient time has elapsed that reconnection has run its course and an island has ``opened.''
In Sec.~\ref{sec:HKgen} we review the Hahm--Kulsrud (HK) solutions and show that their full current sheet is one of a continuous infinity of full current sheet solutions differing by the strength of a constant intensity of current in the sheet, the HK solution ($\tilde\psi_{\rm I}$ in Sec.~\ref{sec:HKgen}) being the one with zero net current.
 
\begin{figure}[htbp]
   \centering
		\includegraphics[width = 0.45\textwidth]{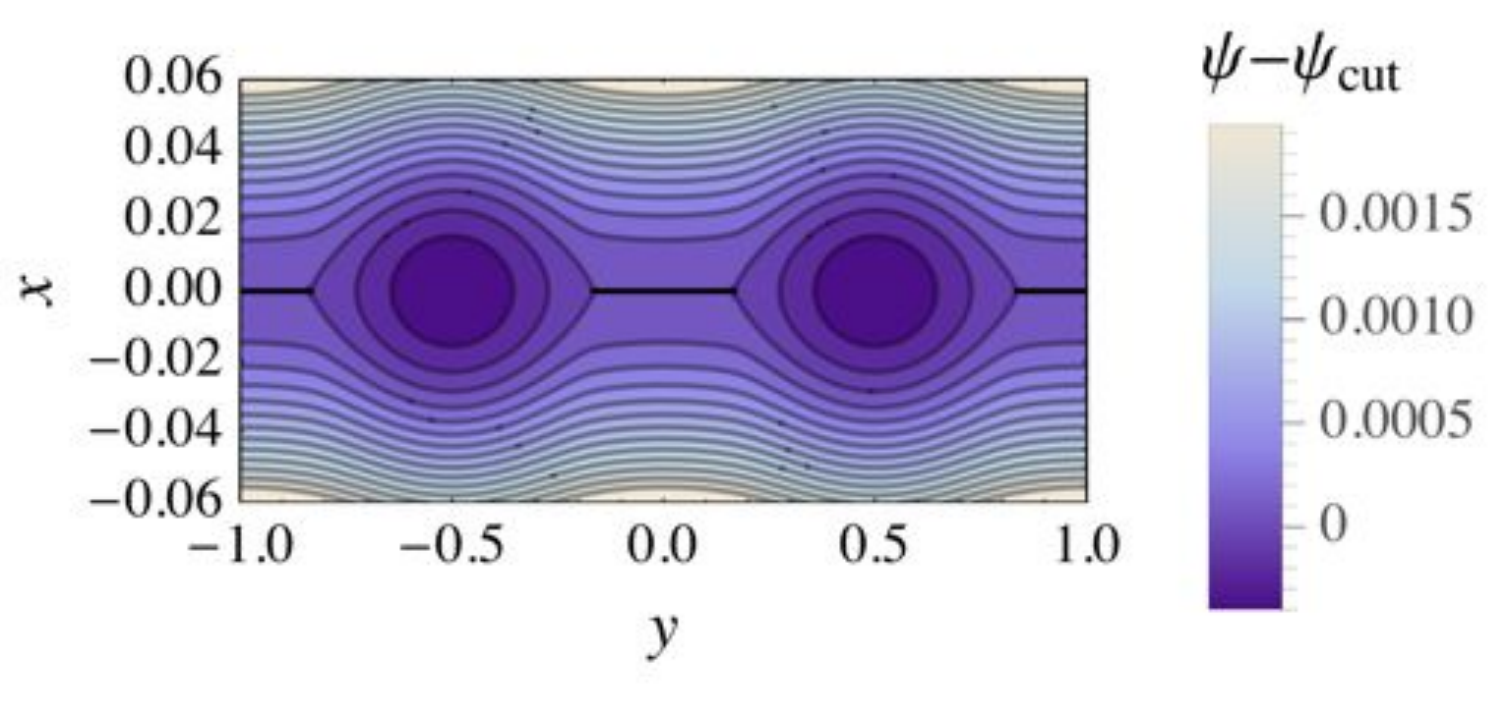} 
\caption{Contours of $\psi - \psi_{\rm cut}$, for the partially-reconnected plasmoid case $L = \lambda/6$ with $\gamma_{\rm S}=1$. Here $\psi_{\rm cut} = -0.000436$ is the value of $\psi$ on the cuts shown as black lines on the $y$-axis. (Color online.)}
\label{fig:TwoThirdsReconnected}
\end{figure}

\begin{figure}[htbp]
   \centering
		\includegraphics[width = 0.35\textwidth]{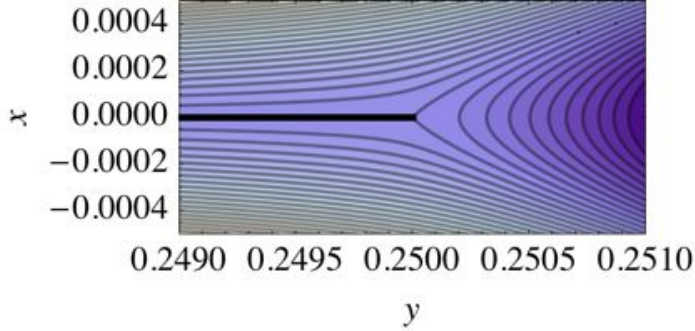} 
\caption{A magnified view of the $\psi$ contours in the vicinity of the junction of a current sheet and a plasmoid, for the case $L = \lambda/4$, $\gamma_{\rm S}=1$, showing the Y-point structure obtained for this value of $\gamma_{\rm S}$ (axes scales equal).}
\label{fig:Zoom}
\end{figure}

The existence of solutions to the HKT equilibrium problem with either a full current sheet or no current sheet raises the possibility that there may be more solutions, intermediate between the two solutions found in Ref.~\onlinecite{Hahm_Kulsrud_85}. As discussed above, current sheets can be unstable to the formation of plasmoids embedded in the current sheet. Thus we seek ideal equilibria that illustrate topologically a scenario for the decay of the shielding HKT current sheet via a plasmoid mechanism of island formation, continuously connecting the two extreme HK solutions. While our solution is not exact, except for the two HK limiting cases, it satisfies \Eq{eq:bndcond} very accurately [error $O(e^{-3 ka})$] so could presumably be made exact by a small perturbation of our ansatz. A typical plasmoid case is depicted in Fig.~\ref{fig:TwoThirdsReconnected}, where current sheets, the black horizontal lines of length $2L = \lambda/3$, alternate with plasmoids of width $2\lambda/3$. A magnified view of a typical current sheet end-point for the case $L=\lambda/2$ is shown in Fig.~\ref{fig:Zoom}.

In Sec.~\ref{sec:Syrovatsky} we review Syrovatsky's complex variable approach to finding an analytic solution of Laplace's equation that represents a large-scale view of a Sweet--Parker current sheet \cite[e.g.]{Yamada_Kulsrud_Ji_10} (see Fig.~\ref{fig:SyrovatskyOverlay}). Syrovatsky's solution was obtained by using a conformal mapping from the simpler solution for the field around a neutral point in the poloidal field.

In Sec.~\ref{sec:PerConformal} we introduce a new, periodic conformal mapping to transform the Syrovatsky solution into a plasmoid solution of the HKT equilibrium problem.

In Sec.~\ref{sec:bnderr} we analyze the mismatch between the boundary condition \Eq{eq:bndcond} and $\psi$ obtained from our conformal mapping ansatz. We present this error both graphically, \emph{vs.} $y$ and $L$ in a typical case, and also give an analytic expression for the first nonvanishing term (6th order!) in an expansion in the small parameter $\epsilon \equiv \exp(-a k/2)$. Possible further improvements and applications of the plasmoid scenario are discussed in Sec.~\ref{sec:conclude}.

\section{Grad--Shafranov equation with current sheets}
\label{sec:GS}

For analyzing equilibria with ideal (zero thickness) current sheets, the force-balance condition is best written in the conservation form  
\begin{equation} \label{eq:weakeqm}
   \divv \left[\left(p  + \frac{B^2}{2 \mu_0}\right)\vsf{I}
     - \frac{\vrm{B B}}{\mu_0} \right] = 0\;,
\end{equation}
where $p$ is the plasma pressure and $\vrm{B}$ is the magnetic field (SI units). In regions where $p$ and $\vrm{B}$ are differentiable, this implies the force balance condition, $\vrm{j}\cross\vrm{B} = \grad p$, where $\vrm{j} = \mu_0^{-1} (\curl\vrm{B}) $ is the plasma current.
However, in the neighborhood of an ideal current sheet, $p$ and $\vrm{B}$ are \emph{not} everywhere differentiable and we need to use generalized functions, like the Dirac delta function $\delta(\varsigma)$, to find weak solutions of \Eq{eq:weakeqm}.

However, we can avoid using generalized functions explicitly by cutting the $x,y$ plane along its intersections with current sheets and solving on the cut plane with appropriate boundary conditions on the cuts.
The boundary conditions on the two sides $\pm$ of a current sheet are found to be \cite[Appendix A]{McGann_Hudson_Dewar_vonNessi_10} the \emph{tangentiality conditions}
\begin{equation} \label{eq:tangcond}
	\vrm{n}\dotv\vrm{B}_{\pm} = 0 \;, \\
\end{equation}
and the \emph{pressure-balance jump condition},
\begin{equation} \label{eq:eqmjumps}
	\jump{p  + \frac{B^2}{2 \mu_0}} = 0 \;.
\end{equation}
The first condition implies that an equilibrium current sheet must be a \emph{tangential discontinuity} in $\vrm{B}$.

In the case of a cylindrical or slab plasma of arbitrary cross section (independent of $z$), a general representation for the equilibrium magnetic field is
\begin{equation}\label{eq:Brep}
	\vrm{B} = \grad z \cross\grad \psi + F (\psi) \grad z \;, 
\end{equation}
where $\psi(x,y)$ is the flux function defined in Sec.~\ref{sec:Intro}, $x,y,z$ being Cartesian coordinates with the $z$-axis in the symmetry direction. The first of Eqs.~(\ref{eq:eqmjumps}) implies that the two sides of an equilibrium current sheet are level surfaces of $\psi_{\pm}$ (in fact $\psi$ must be continuous across the current sheet, $\psi_{-}=\psi_{+}$, to avoid infinite poloidal magnetic field there) while the second gives
\begin{equation} \label{eq:2Djumpcond}
	\jump{p  + \frac{|\grad\psi|^2 + F^2}{2 \mu_0}} = 0 \;,
\end{equation}

Taking the curl of \Eq{eq:Brep} we find, everywhere except on a cut,
\begin{equation}\label{eq:jrep}
	\mu_0\vrm{j} = \nabla^2\psi\, \grad z + F'(\psi)\grad\psi\cross\grad z \; ,
\end{equation}
where $\nabla^2$ is the 2-dimensional Laplacian, $\partial_x^2 + \partial_y^2$, and $F' \equiv \partial F/\partial\psi$.

Summarizing, the equilibrium condition \Eq{eq:weakeqm} is satisfied if and only if the two-dimensional \emph{Grad--Shafranov} equation for axisymmetric static MHD equilibria,
\begin{equation}\label{eq:GSpointwise}
	\nabla^2 \psi + \partial_\psi\left[\mu_0 p (\psi) + \half F (\psi)^2\right] = 0 \;,
\end{equation}
is satisfied everywhere except on cuts, where the current sheet force-balance condition \Eq{eq:2Djumpcond} applies instead. 

Equation (\ref{eq:GSpointwise}) is in general nonlinear, but consider the special case $\mu_0 p (\psi) + \frac{1}{2} F (\psi)^2 = {\rm const}\, - (B_a/a)\psi$ for which \Eq{eq:GSpointwise} becomes a Poisson equation, $\nabla^2 \psi  = B_a/a$, linear in $\psi$. This can be solved as a linear superposition, $\psi_{\rm 0} + \psi_{\rm h}$ where $\psi_{\rm h}$ is a \emph{harmonic function}, i.e. a solution of the Laplace equation
\begin{equation}\label{eq:LaplaceEq}
	\nabla^2\psi_{\rm h} = 0 \;,
\end{equation}
determined by the boundary conditions. Comparing with \Eq{eq:psitildedef} we identify $\psi_{\rm h}$ with $\alpha\tilde\psi$. The wall boundary conditions \Eq{eq:bndcond}, current sheets on the $y$-axis, the assumed form of $\psi_0(x)$, and \Eq{eq:LaplaceEq}, make up what we call the \emph{HKT equilibrium problem}, whose scope we expand by considering a wider class of current sheet cuts in the domain on which \Eq{eq:LaplaceEq} is to be solved. This has the consequence that $\tilde\psi$ cannot be assumed necessarily to be sinusoidal in $y$.

Noting that $\jump{\psi} = 0$ we see from the assumption $\mu_0 p (\psi) + \frac{1}{2} F (\psi)^2 = {\rm const} - (B_a/a)\psi$ that the equilibrium jump condition \Eq{eq:2Djumpcond} simplifies to
\begin{equation} \label{eq:reducedjumpcond}
	\jump{|\grad\psi|^2} = 0 \;.
\end{equation}
Although $\psi$ is continuous, and by \Eq{eq:reducedjumpcond}, $|\grad\psi|$ is continuous, $\grad\psi$ can be discontinuous, its jump giving the intensity $j_{\ast}$ of the $\delta$-function component of the current $\vrm{j} = \curl\vrm{B}/\mu_0$,
\begin{multline} \label{eq:sheetcurrent}
	 \phantom{\quad\quad} \vrm{j} = \frac{1}{\mu_0}\left[F'(\psi)\grad\psi\cross\grad z \phantom{\frac{B_a}{a}}
	\right. \\ \left. + \left(\frac{B_a}{a} + \jump{\partial_x\psi}\delta(x)\right)\!\grad z \right] \phantom{\quad\quad}
\end{multline}
at each point on the current sheets on the $y$-axis, whence $j_{\ast}(y) = \jump{\partial_x\psi}/\mu_0$.

\begin{figure}[hbtp]
\begin{center}
\includegraphics[width=0.45\textwidth]{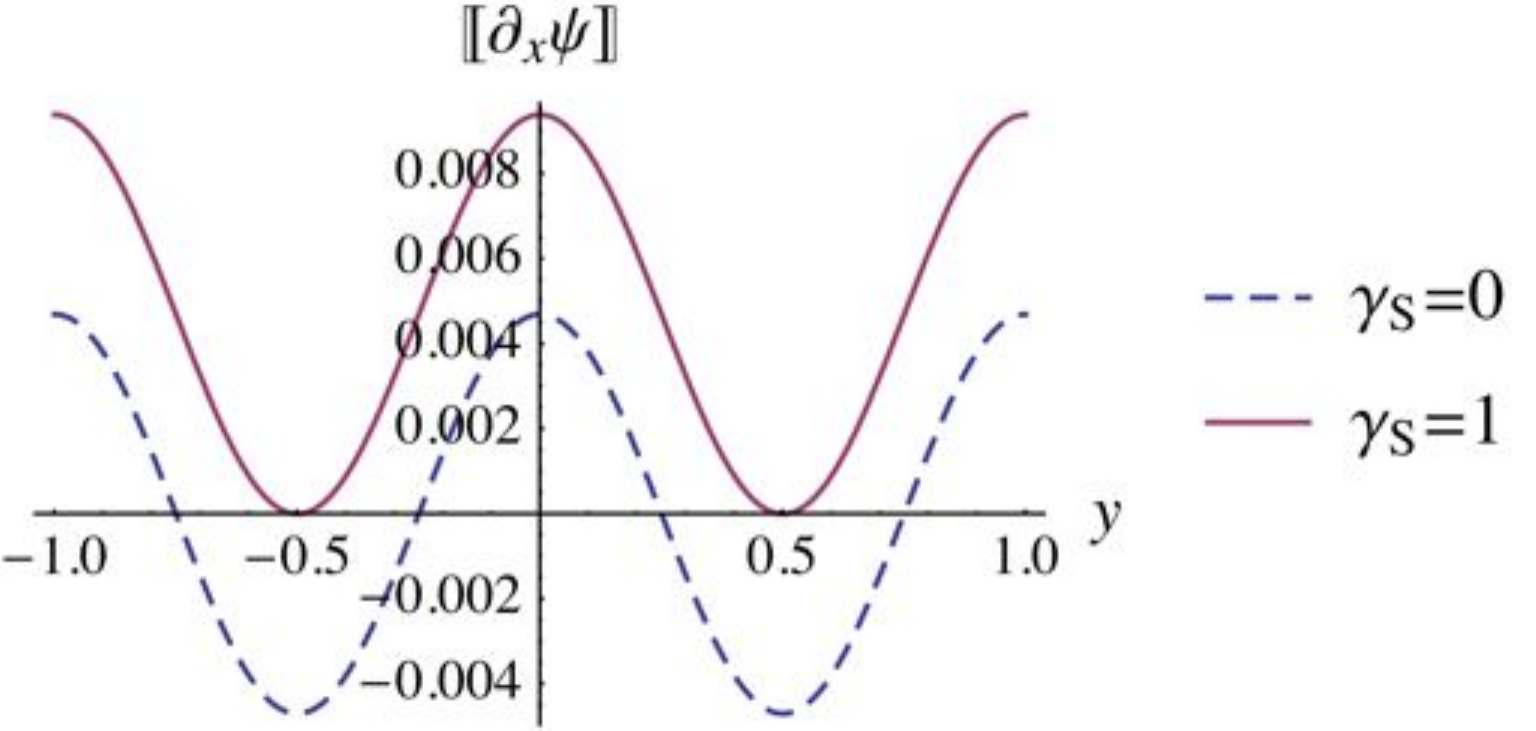}
\caption{Jumps in the gradient of $\psi$ across the $y$-axis, giving the intensity of the sheet currents by \Eq{eq:sheetcurrent}, for the original HK fully screened solution $\gamma_{\rm S}=0$ (dashed) and the modified HK solution $\gamma_{\rm S}=1$ (solid) for which the current sheet intensity never goes negative.}
\label{fig:HKjump}
\end{center}
\end{figure}

\section{Generalized HK shielding solutions}
\label{sec:HKgen}

Hahm and Kulsrud \cite{Hahm_Kulsrud_85} found two solutions of the HKT equilibrium problem, a \emph{shielding current sheet} solution $\tilde\psi_{\rm I}(x,y) = a B_a|\!\sinh(kx)|\cos(ky)/\!\sinh(ka)$, and a fully developed \emph{island} solution $\tilde\psi_{\rm II}(x,y) = a B_a\cosh(kx)\cos(ky)/\!\cosh(ka)$ with no current sheet \footnote{Hahm and Kulsrud also considered the general solution $\tilde\psi = A\psi_{\rm II}+ (1-A)\psi_{\rm I}$, taking the constant $A(t)$ [expressed in terms of the reconnected flux $\alpha \psi_{\rm II}(0,0) A(t)$] to represent the time evolution of the reconnection process.}

\begin{figure}[htbp]
\begin{center}
\includegraphics[width=0.4\textwidth]{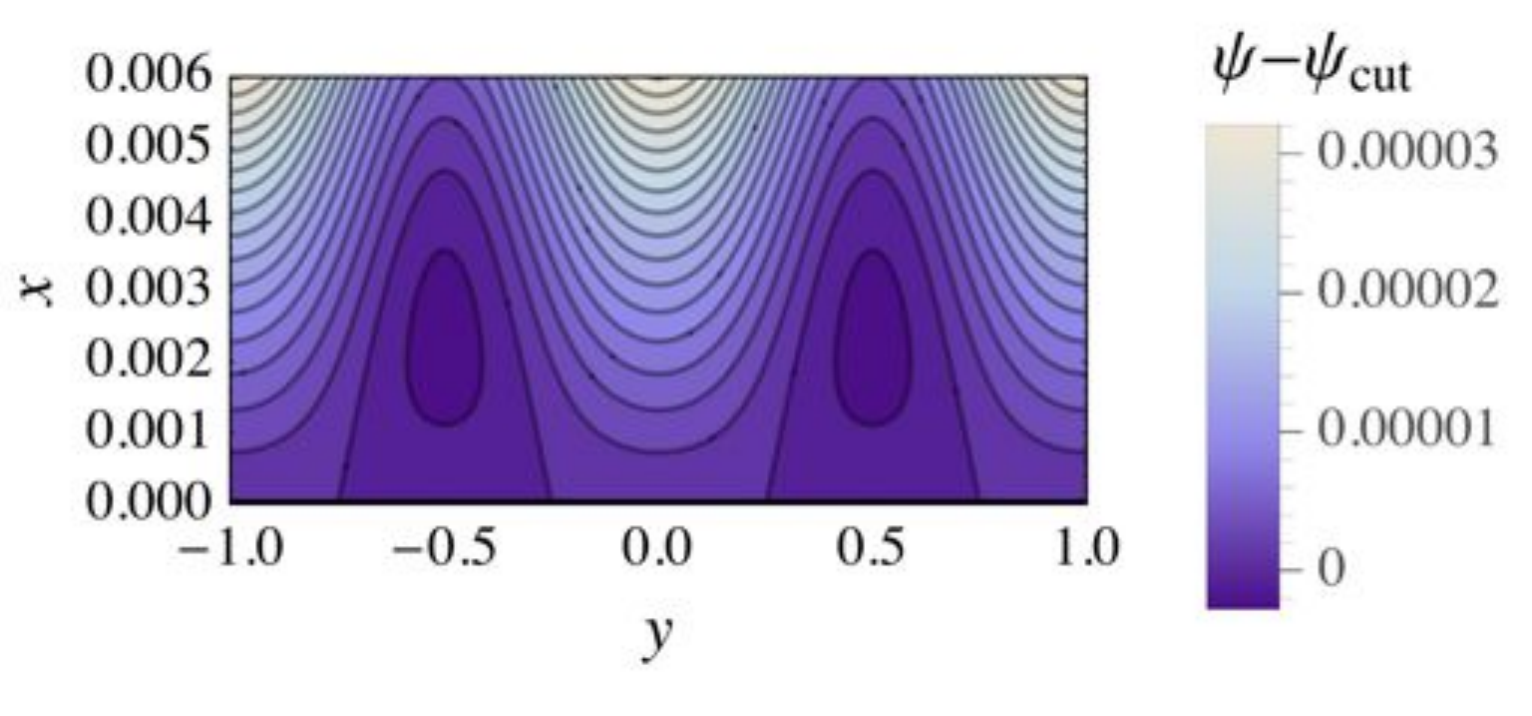}
\caption{Contours of $\psi - \psi_{\rm cut}$ near the $y$-axis for the original HK fully screened solution $\psi = \psi_0 + \alpha\psi_{\rm I}$ ($\psi_{\rm I^*}$ for $\gamma_{\rm S}=0$), showing islands with separatrices joining points where the sheet current changes sign.}
\label{fig:HKcont}
\end{center}
\end{figure}

\begin{figure}[htbp]
\begin{center}
\includegraphics[width=0.4\textwidth]{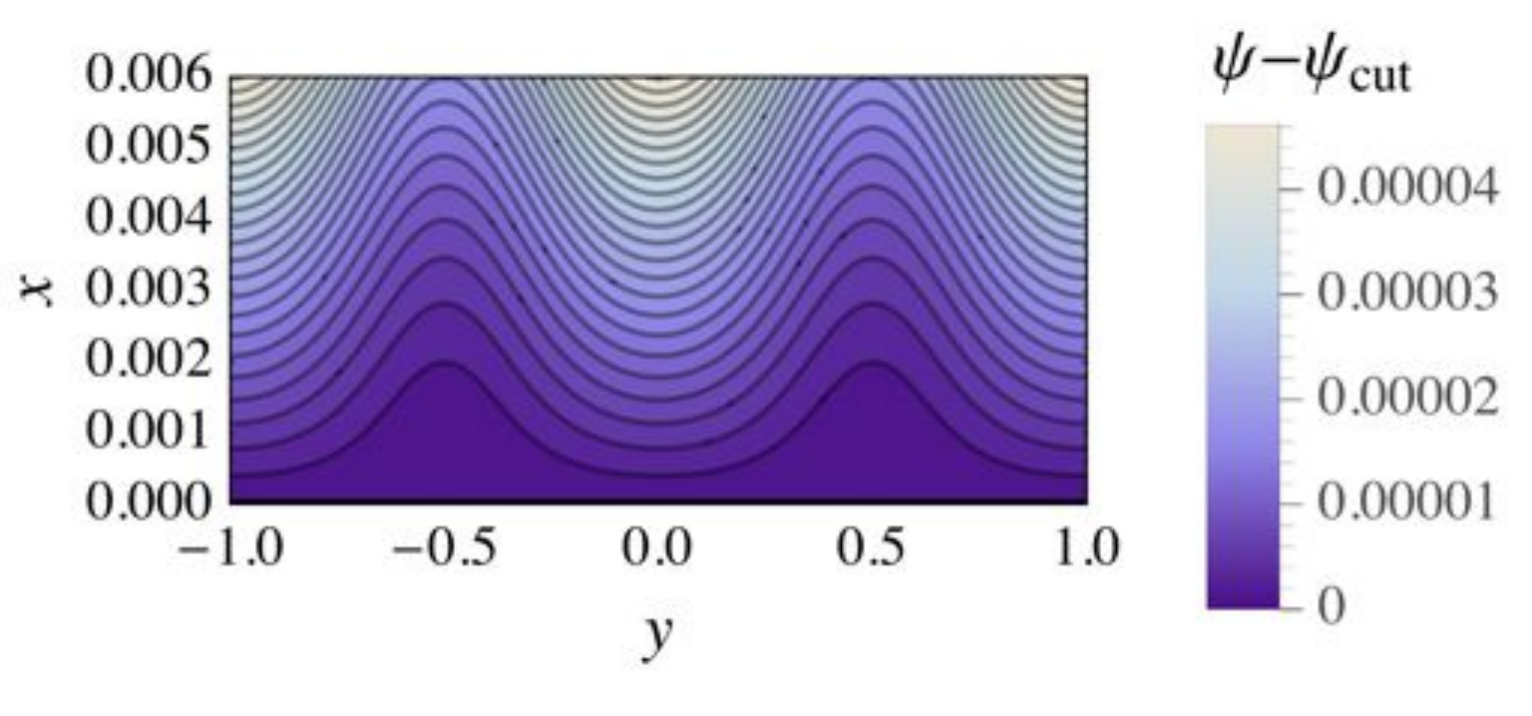}
\caption{Contours of $\psi - \psi_{\rm cut}$ near the $y$-axis for the modified fully screened solution $\psi = \psi_0 + \alpha\psi_{\rm I^*}$ with $\gamma_{\rm S}=1$, showing the simple topology of the magnetic surfaces in this case.}
\label{fig:HKgencont}
\end{center}
\end{figure}

It is easily verified that
\begin{equation}\label{eq:HKgen}
	 \tilde\psi_{\rm I^*}(x,y) = a B_a\frac{ |\!\sinh (k x)|\cos (k y) + \gamma_{\rm S}  k (|x| - a)}{\sinh(ka)}\;,
\end{equation}
where $\gamma_{\rm S}$ is an arbitrary constant, also satisfies the HKT equilibrium problem. The inclusion of the $\gamma_{\rm S}$ term represents a small but important generalization of the HK shielding current sheet solution that allows the dc level of the current in the sheet to be adjusted, as illustrated in Fig.~\ref{fig:HKjump}.

This has a rather profound influence on the topology of the $\psi$ contours, as illustrated in Figs.~\ref{fig:HKcont} and \ref{fig:HKgencont} where it is seen that the HK solution $\psi$ has saddle points at the current-reversal points $x = 0,\: y = \lambda/4 \pm n\lambda/2$, $n = 0, 1, 2, \ldots$, where the $\psi = 0$ contour bifurcates off the $y$-axis, forming magnetic islands.

In the following sections we obtain these solutions as limiting cases of a new family of solutions $\tilde\psi$ obtained using a conformal mapping approach \cite{Milne-Thomson_68}, which relies on the facts that the real or imaginary part of \emph{any} analytic function, $F_{\rm S}(\zeta)$, $\zeta \equiv x +iy$, is harmonic, and that the composition of two analytic functions is itself analytic.

\section{The Syrovatsky current sheet}
\label{sec:Syrovatsky}

The double-valued analytic functions \cite{Syrovatsky_71,Parker_Dewar_Johnson_90}
\begin{align}
\label{eq:Syrovatsky}
    F_{\rm S}(u) &=  \sgn(\Re u)\left\{u (u^2  + 1)^{1/2} \right. \notag \\
    & \quad\quad\quad\quad+ \left. \gamma_{\rm S}\ln[u+ (u^2 + 1)^{1/2}]\right\} \;, \\
\label{eq:Syrovatskyprime}
    F'_{\rm S}(u) &= \sgn(\Re u)\frac{2 u^2 + 1 + \gamma_{\rm S}}{(u^2 + 1)^{1/2}}  \;,
\end{align}
defined on the complex $u$-plane with two Riemann sheets joined by a cut joining branch points $u = \pm i$, may be used to define the harmonic functions $\psi_{\rm S}(x',y') \equiv \Re F_{\rm S}(x' + iy')$ and $\partial_{x'}\psi_{\rm S}(x',y') = \Re F'_{\rm S}(x' + iy')$. The former can be interpreted as the flux function for a Sweet--Parker current sheet positioned on the cut between $y' = \pm 1$ and the latter gives the intensity of the current sheet on the cut.

The form in \Eq{eq:Syrovatsky} is that used in Ref.~\onlinecite{Parker_Dewar_Johnson_90}, with square root $z^{1/2}$ and natural logarithm $\ln z$ defined as usual on the complex $z$-plane cut along the negative real axis. The step function factor, $\sgn(\Re u) = \pm 1$, is needed to make the cut a straight line joining $y' = \pm 1$ rather than make two cuts radiating outward to infinity. The function defined by \Eq{eq:Syrovatsky} is obtained by the following substitutions in Syrovatsky's \cite{Syrovatsky_71} Eq.~(44),
$$n\mapsto 2, b\mapsto i, A\mapsto \frac{\pi  i}{2}, \alpha \mapsto 2, \Gamma \mapsto 2 \pi\gamma_{\rm S}, z\mapsto \zeta'\;.$$

It is seen from \Eq{eq:Syrovatskyprime} that in general the sheet current intensity diverges at the endpoints $y' = \pm 1$, but the choice $\gamma_{\rm S} = 1$ \cite[Eq.~(47)]{Syrovatsky_71} makes the current go continuously to zero at the endpoints, giving rise to the Y-points seen in Fig.~\ref{fig:SyrovatskyOverlay}. This figure shows contours of $\psi_{\rm S}(x',y')$ in the $y',x'$ plane, the thick horizontal line indicating the cut/current sheet and the thinner continuous lines the magnetic field lines. The ellipses and hyperbolae form a visualization of a periodic conformal map shown in the next section to convert Syrovatsky's single current sheet solution into a plasmoid solution of the HKT equilibrium problem.

\begin{figure}[hbtp]
\begin{center}
\includegraphics[width=0.45\textwidth]{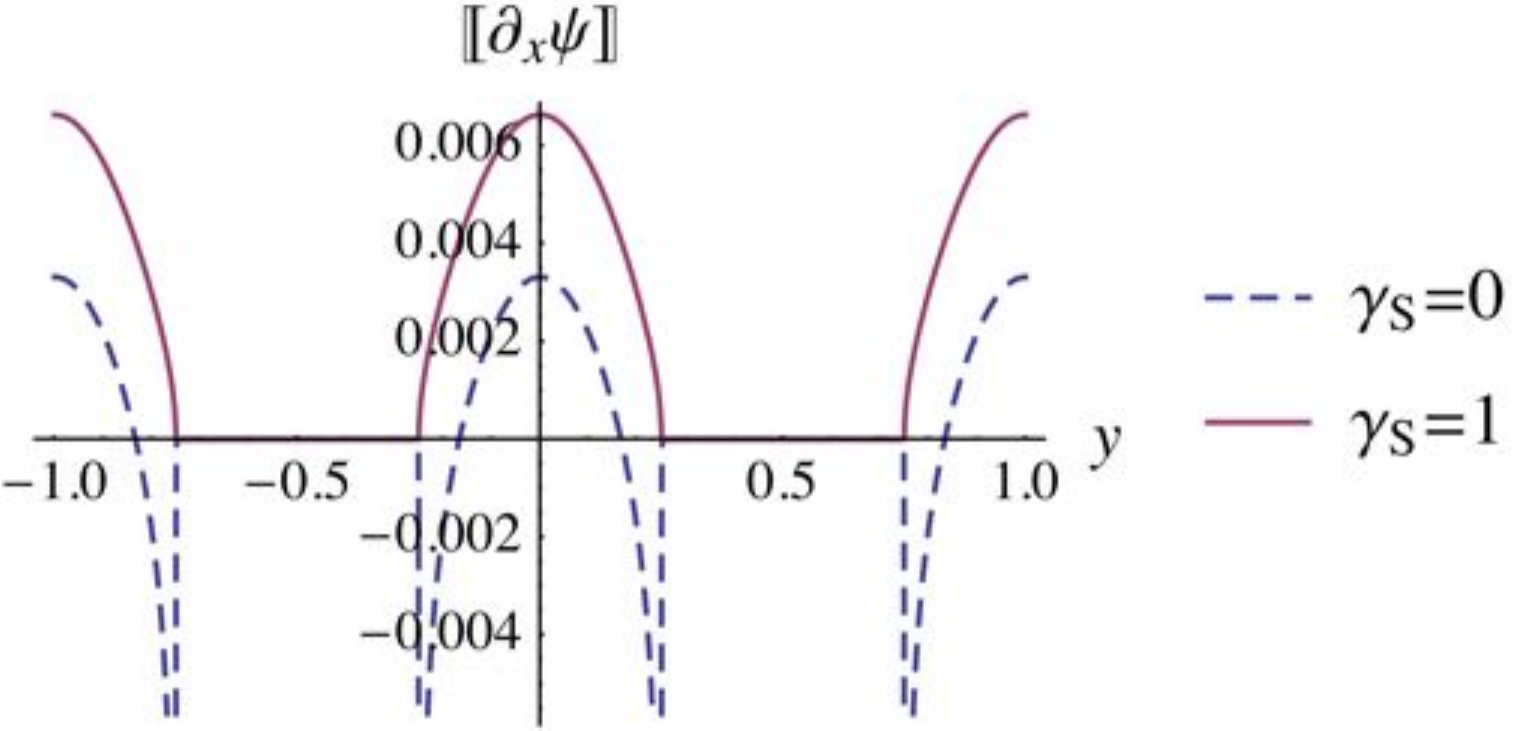}
\caption{Jumps in the gradient of $\psi$ across the $y$-axis for the half-screened plasmoid case $L = \lambda/4$. The dashed curve is for the case $\gamma_{\rm S}=0$, showing current reversal within the current sheets and singularities at the ends, and the solid curve is for the case $\gamma_{\rm S}=1$ in which the current sheet intensity never goes negative and joins continuously to the zero values within the plasmoid regions.}
\label{fig:HalfReconnectedJump}
\end{center}
\end{figure}

\section{Shinusoidal transformation}
\label{sec:PerConformal}

The analytic function
\begin{equation}
	f(\zeta) \equiv \frac{\sinh (\half k \zeta)}{\sin(\half kL)}
	\label{eq:confmap}
\end{equation}
will be used to map the single current sheet in the Syrovatsky solution \Eq{eq:Syrovatsky} to a periodic sequence of current sheets of length $2L$, replacing the X-points at $y =  n\lambda$, $n = 0, \pm1, \pm 2, \ldots$, in the fully reconnected (magnetic island) solution of Ref.~\onlinecite{Hahm_Kulsrud_85}, where $\lambda$ is the wavelength of the boundary perturbation. 

\begin{figure}[htbp]
\begin{center}
\includegraphics[width=0.4\textwidth]{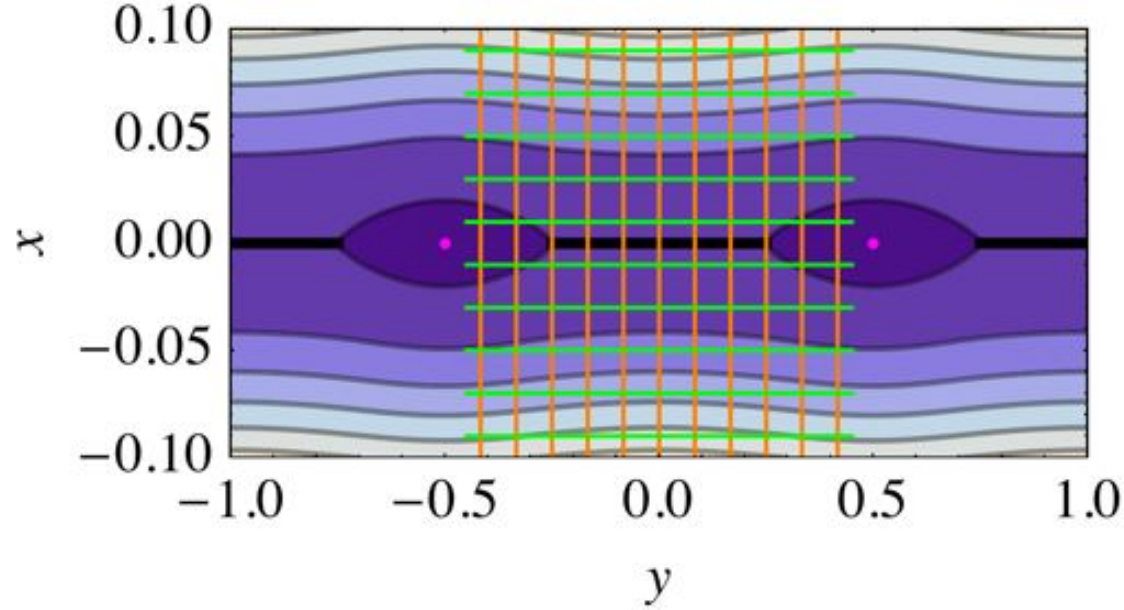}
\caption{Contours of $\psi = \psi_0+\alpha\tilde\psi$ for a plasmoid case $L = \lambda/4$, $\gamma_{\rm S} = 1$, with magnified scale on $x$-axis. Mapped images of the vertical (orange) and horizontal (green) mesh lines appear in Fig.~\ref{fig:SyrovatskyOverlay} (color online).} 
\label{fig:HalfHKTcontours}
\end{center}
\end{figure}

Its appropriateness for this purpose will be verified below \emph{a posteriori}, but as partial motivation for this ansatz we note some useful properties of $f$:
\begin{itemize}
	\item $f(x) = \sinh(\half kx)/\sin(\half kL)$\\
	$f(iy) = i\sin(\half ky)/\sin(\half kL)$\\
	$f(x \pm \half i\lambda) =\pm i\cosh(\half kx)/\sin(\half kL)$
	\item $f(x+iy)$ is periodic in $y$ with wavelength $2\lambda$, but $f(x+iy)^2$ has wavelength $\lambda$.
	\item The only zeros of $f(\zeta)$ are at $\zeta = i n\lambda$, i.e. at the X-points of the fully reconnected solution, while at the first O-point, $y=\lambda/2$, $f(i\lambda/2) = i/\sin(\half kL)$ ranges from $i$ in the fully reconnected case, $L=\lambda/2$, to $i\infty$ in the fully shielded case, $L=0$.
	\item The double-valued function $\sin(\half kL)[1 + f(y)^2]^{1/2} \equiv [(\sin\half kL-\sin\half k\zeta)(\sin\half kL+\sin\half k\zeta)]^{1/2}$ has, within the strip $-\lambda/2 < \Im\zeta \leq \lambda/2$, two branch points, located at the endpoints $\zeta = \pm L$, of the sought-for current sheet,.
\end{itemize}

\begin{figure}[htbp]
\begin{center}
\includegraphics[width=0.4\textwidth]{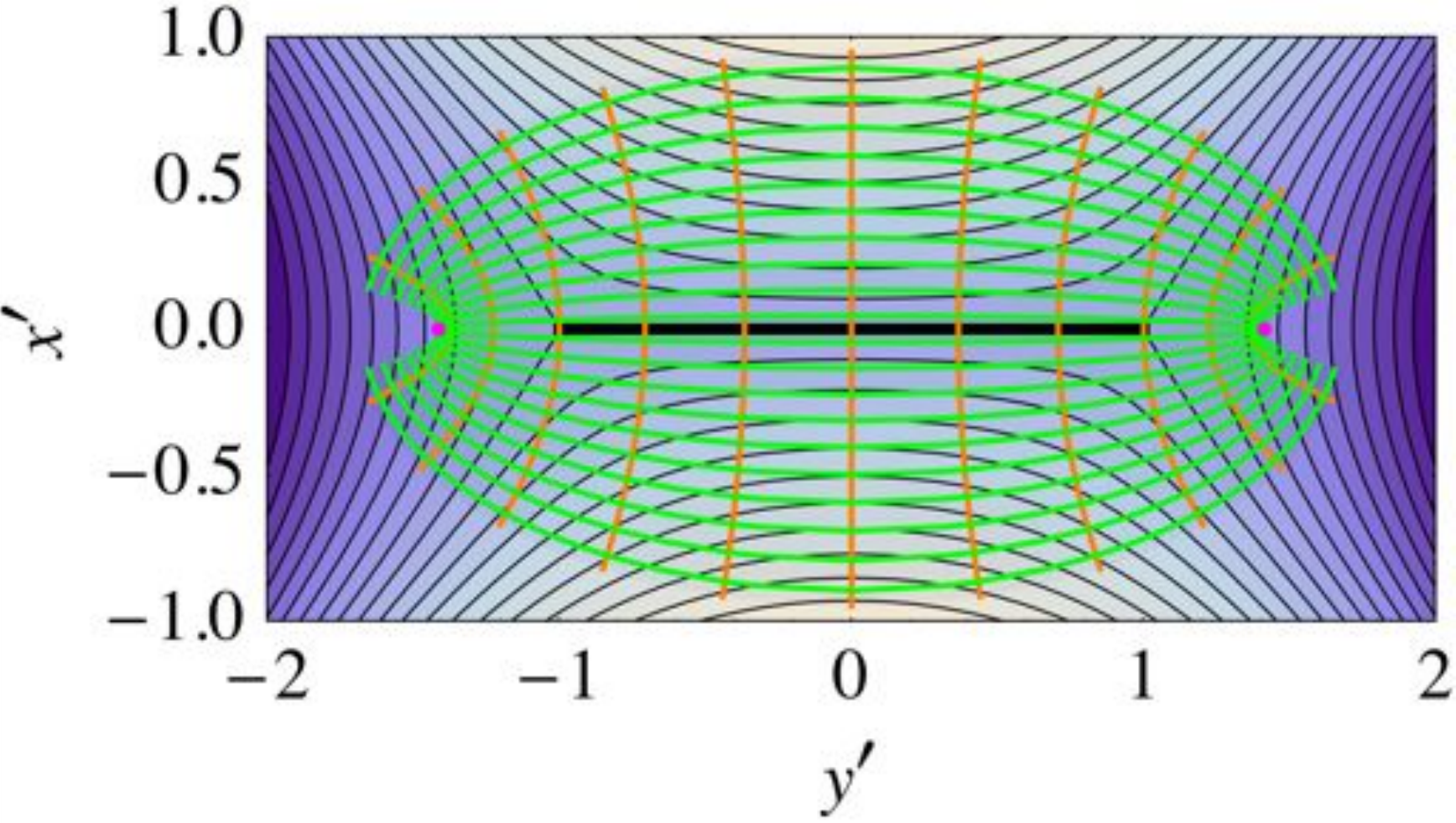}
\caption{Contours (black) in the $y',x'$-plane of the Syrovatsky solution $\psi_{\rm S}(x',y')=\Re F_{\rm S}(x',y')$ given in Sec.~\ref{sec:Syrovatsky} in the case $\gamma_{\rm S} = 1$. The orange hyperbolae and green ellipses are the images of the rectangular mesh in Fig.~\ref{fig:HalfHKTcontours}, extended to the range $x \in [-0.2,0.2]$, under the periodic conformal map $x'+iy'=f(x+iy)$ defined in Sec.~\ref{sec:PerConformal} (axes scales equal; color online).} 
\label{fig:SyrovatskyOverlay}
\end{center}
\end{figure}

We now use the conformal map $\zeta' = f(\zeta)$ to transform the Syrovatsky function to a function of $\zeta$, $F_{\rm S}(\zeta^{\prime}) = F_{\rm S}\!\!\circ\!\! f(\zeta)$, that provides the harmonic function $\tilde\psi(x,y)$ through the equation
\begin{equation}
	\tilde\psi(x,y) = aB_a\,\frac{c + \Re F_{\rm S}(\zeta')}{d} \;,
	\label{eq:tildepsisoln}
\end{equation}
where $\zeta' = f(x+iy)$ and the constants $c$ and $d$ are determined by requiring that the boundary condition $\tilde\psi(\pm a,y) = aB_0\cos ky$ be satisfied to a good approximation (see Sec~\ref{sec:bnderr}).

Figure~\ref{fig:TwoThirdsReconnected} illustrates how this transformation results in a typical plasmoid structure for $0 < L < \lambda/2$. A graphical visualization of the transformation may be had by comparing the mesh in Fig.~\ref{fig:HalfHKTcontours}  with its image in Fig.~\ref{fig:SyrovatskyOverlay}. Figure~\ref{fig:Boundary} illustrates the decay of the ripple away from the boundary before it is amplified by the resonance effect near the poloidal field reversal region, $|x| \sim 0$, as seen in Figs.~\ref{fig:HKcont} and \ref{fig:HKgencont}.
 
Figure~\ref{fig:HalfReconnectedJump} shows the current sheet intensity for the cases case $\gamma_{\rm S}=1$ and case $\gamma_{\rm S}=0$, showing the singular behavior inherited from the Syrovatsky solution \Eq{eq:Syrovatskyprime} in the latter case. Figure~\ref{fig:Zoom} verifies that the case $\gamma_{\rm S} = 1$ leads to the typical Y-point magnetic surface behavior expected of a Sweet--Parker current sheet, whereas Syrovatsky \cite[Fig.~3]{Syrovatsky_71} showed that the case $\gamma_{\rm S} = 0$ leads to reentrant  $\psi$-contours producing a cusp pointing away from the current sheet.

It may be shown analytically that the solution \Eq{eq:tildepsisoln} reduces to the HK island solution $\tilde\psi(x,y) = a B_a\cosh(kx)\cos(ky)/\!\cosh(ka)$ as $L \to 0$, while it reduces to the generalized HK current sheet solution \Eq{eq:HKgen} as $L \to \lambda/2$.

\begin{figure}[htbp]
\begin{center}
\includegraphics[width=0.35\textwidth]{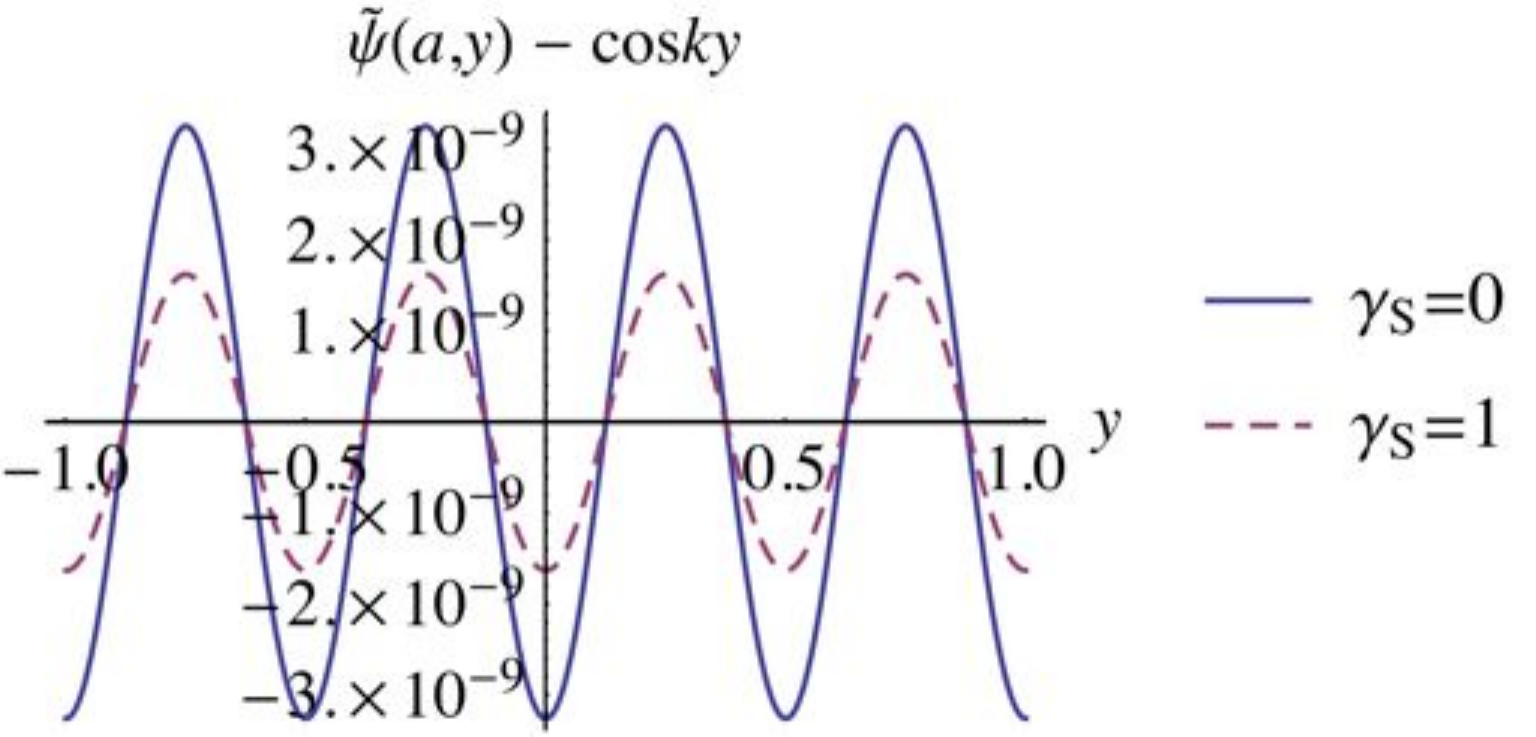}
\caption{The residual boundary error $\tilde\psi(a,y) - \cos ky$ \emph{vs.} $y$ for the case $L = \lambda/4$.}
\label{fig:BE}
\end{center}
\end{figure}

\begin{figure}[htbp]
\begin{center}
\includegraphics[width=0.35\textwidth]{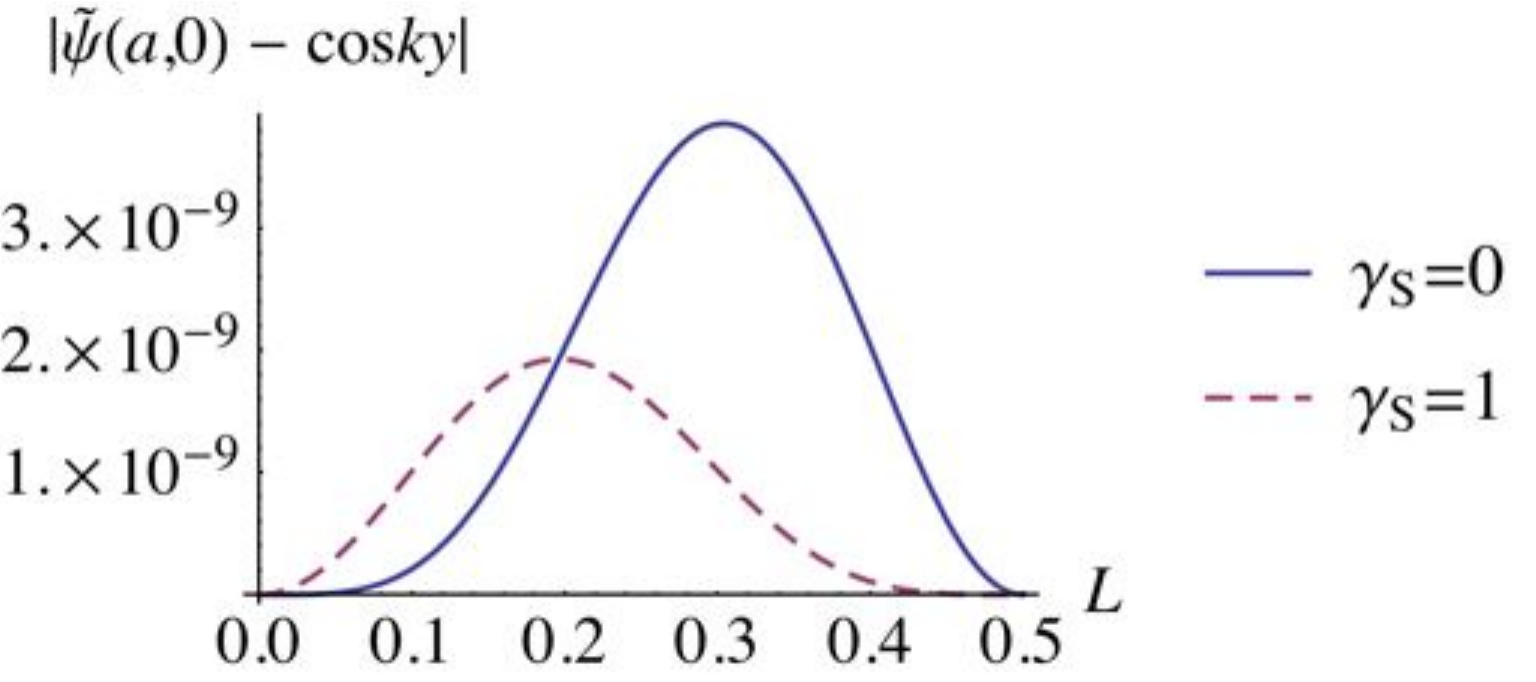}
\caption{The residual boundary error maximized over $y$, i.e. $|\tilde\psi(a,0) - 1|$, \emph{vs.} $L$.}
\label{fig:BEvsL}
\end{center}
\end{figure}

\section{Boundary error analysis}
\label{sec:bnderr}

The current sheet force-balance requirement \Eq{eq:reducedjumpcond} is ensured by the restriction of the cuts to the $y$-axis, and the assumed symmetry about this axis, but the boundary conditions \Eq{eq:bndcond} are not imposed \emph{a priori} for all $y$. Instead we impose only two conditions involving the \emph{boundary error function},
\begin{equation}
	\varepsilon_a(y) \equiv c + \Re F_{\rm S}[f(a + iy)]  - d\cos ky \;,
	\label{eq:bnderrdef}
\end{equation}
in order to determine the two constants $c$ and $d$ in \Eq{eq:tildepsisoln}. The two conditions are
\begin{equation}
	\varepsilon_a(0) - \varepsilon_a(\lambda/2) = 0 \:\:\text{and}\:\: \varepsilon_a(0) + \varepsilon_a(\lambda/4) = 0 \;.
	\label{eq:cdeqns}
\end{equation}

We can now verify \emph{a posteriori} that the boundary conditions are satisfied to high accuracy for all $y$ in typical cases. For instance, in Fig.~\ref{fig:BE} we plot $\varepsilon_a(y)$ for the same case as shown in Fig.~\ref{fig:HalfHKTcontours} and see that the conditions in \Eq{eq:cdeqns} null out any constant error and the fundamental, $\cos ky$, leaving only a second harmonic error proportional to $\cos 2ky$, with an amplitude that is extremely small in the case studied.

Noting that the error is an even function, periodic in $y$, we see that $y=0$ is a maximum point of the absolute value of the error. In Fig.~\ref{fig:BEvsL} we plot this maximum error \emph{vs.} the halfwidth, $L$, of the current sheet. This figure shows that the error is \emph{zero} for the two HKT cases $L=0$ (complete reconnection) and $L=\lambda/2$ (complete shielding), and nowhere gets much larger than it does in the typical intermediate case depicted in Figs.~\ref{fig:HalfHKTcontours} and \ref{fig:BE}.

In the plots Eqs.~\ref{eq:cdeqns} have been solved numerically, but to understand why the error is so extraordinarily small it is instructive to perform a perturbation expansion in $\epsilon \equiv \exp(-a k/2)$ ($= 0.0432\ldots$ in the reference case used in this paper), which allows us to separate those terms that are large at $x=a$ from those that are small. For instance,
\begin{equation}
	f(a+iy) \equiv \frac{\epsilon^{-1}\exp(\half iky) - \epsilon\exp(-\half iky)}{2\sin(\half kL)} \;.
	\label{eq:confmapasymp}
\end{equation}

Expanding \Eq{eq:tildepsisoln} in $\epsilon$ and solving for $c$ and $d$ such that the constant and $\cos ky$ terms in $\varepsilon_a(y)$ vanish we find, to first nonvanishing order, the residual boundary error
\begin{multline}
	\tilde\psi(a,y) - aB_a\cos ky =  -\frac{aB_a}{4} e^{-3 a k} \sin ^2 k L  \\
	\times\left[2-\gamma_{\rm S} - (2 - 3 \gamma_{\rm S}) \cos k L\right] \cos 2 k y
	\label{eq:residerrasymp}
\end{multline}
which is $O(\epsilon^6)$, thus explaining the smallness of the error found numerically.

\section{Conclusion}
\label{sec:conclude}

This paper demonstrates that solutions to the global ideal-MHD equibrium problem are far from unique when interior current sheets are allowed. We have made no attempt to link the members of our equilibrium sequence by applying constraints consistent with almost-ideal-MHD time evolution, so this work, of itself, cannot be regarded as a study of reconnection. However, it is highly suggestive that evolution through a plasmoid phase represents a topologically reasonable mechanism for an initial shielding current sheet to open into a magnetic island. To establish this scenario as a reconnection mechanism, two approaches appear promising, both applying a subset of the ideal-MHD constraints: 
\begin{enumerate}
\item A \emph{maximally constrained} or \emph{almost-ideal MHD} approach assuming ideal MHD applies locally throughout the evolution, except as a plasma passes through a Sweet--Parker current sheet where the frozen-in-flux constraint is relaxed and reconnection can occur. While respecting the detailed physics of the process, it is not amenable to the conformal mapping approach we have used to find analytical solutions as it does not preserve the condition of linearity of $\mu_0 p(\psi)  + F(\psi)^2/2$ assumed at the beginning of Sec.~\ref{sec:PerConformal}. Furthermore, it implies current sheets on the plasmoid separatrices, so that the simple cut structure of the Syrovatsky solution does not apply \cite{Uzdensky_Kulsrud_97,Wang_Bhattacharjee_95}. Thus a completely different method of analysis would need to be applied. An interesting approach has been discussed by Kulsrud \cite{Kulsrud_11}.

\item A \emph{minimally constrained} or \emph{relaxed MHD} approach based on a generalization of Taylor \cite{Taylor_86} relaxation to include more ideal-MHD invariants than the magnetic helicity constraint assumed by Taylor, but only a sufficient number to capture the qualitative essence of the evolution (cf. \cite{Bhattacharjee_Dewar_82,Hudson_etal_12b}). A noncanonical Hamiltonian approach has recently been developed  \cite{Yoshida_Dewar_12} in which the ideal-MHD constraints appear as Casimir invariants. In this work it was shown that bifurcation of a cylindrical Taylor relaxed state to a helical relaxed state can be frustrated by introducing a singular Casimir invariant corresponding to the shielding HKT current sheet, the magnetic field everywhere else in the plasma being given by the linear Beltrami equation, $\curl\vrm{B} = \mu\vrm{B}$, found by Taylor. This suggests seeking, in slab geometry, a sequence of plasmoid solutions analogous to those found in the present paper, especially in the limit $\mu \to 0$ where the Beltrami field reduces to a harmonic field corresponding to $\tilde\psi$.
\end{enumerate}

Using either approach to generate an equilibrium sequence with fixed boundary conditions, its applicability as a physically plausible reconnection scenario could be determined, without the necessity of resolving the current sheets into finite-width tearing layers, simply by showing that the plasma potential energy $W = \int [p/(\gamma-1) + B^2/2\mu_0]dV$ \cite{Kruskal_Kulsrud_58} decreases monotonically along the sequence, the final state being a minimum of $W$. Presumably, if two sequences are parametrized by their reconnected fluxes and the graph of the potential energy of one lies below that of the other, then the first sequence is physically preferred. This could be used to determine when and if the symmetry-breaking plasmoid evolution found in Ref.~\onlinecite{Parker_Dewar_Johnson_90} can occur, rather than the symmetric evolution assumed in the present paper.

\section*{Acknowledgments}
One of the authors (RLD) would like to thank the hospitality of and stimulating conversations with Roger Hosking, as the idea behind this paper was conceived during work on our book, still in preparation, ``Fundamentals of Fluid Mechanics and MHD.'' He would also like to thank the hospitality of Princeton Plasma Physics Laboratory where the first draft was written and of Zensho Yoshida at the University of Tokyo where the work was completed. This research has been supported by the Australian Research Council and the U.S. National Science Foundation and Department of Energy. The plots were made using Mathematica 9 \cite{Mathematica9}.


\begin{thebibliography}{27}%
\makeatletter
\providecommand \@ifxundefined [1]{%
 \@ifx{#1\undefined}
}%
\providecommand \@ifnum [1]{%
 \ifnum #1\expandafter \@firstoftwo
 \else \expandafter \@secondoftwo
 \fi
}%
\providecommand \@ifx [1]{%
 \ifx #1\expandafter \@firstoftwo
 \else \expandafter \@secondoftwo
 \fi
}%
\providecommand \natexlab [1]{#1}%
\providecommand \enquote  [1]{``#1''}%
\providecommand \bibnamefont  [1]{#1}%
\providecommand \bibfnamefont [1]{#1}%
\providecommand \citenamefont [1]{#1}%
\providecommand \href@noop [0]{\@secondoftwo}%
\providecommand \href [0]{\begingroup \@sanitize@url \@href}%
\providecommand \@href[1]{\@@startlink{#1}\@@href}%
\providecommand \@@href[1]{\endgroup#1\@@endlink}%
\providecommand \@sanitize@url [0]{\catcode `\\12\catcode `\$12\catcode
  `\&12\catcode `\#12\catcode `\^12\catcode `\_12\catcode `\%12\relax}%
\providecommand \@@startlink[1]{}%
\providecommand \@@endlink[0]{}%
\providecommand \url  [0]{\begingroup\@sanitize@url \@url }%
\providecommand \@url [1]{\endgroup\@href {#1}{\urlprefix }}%
\providecommand \urlprefix  [0]{URL }%
\providecommand \Eprint [0]{\href }%
\providecommand \doibase [0]{http://dx.doi.org/}%
\providecommand \selectlanguage [0]{\@gobble}%
\providecommand \bibinfo  [0]{\@secondoftwo}%
\providecommand \bibfield  [0]{\@secondoftwo}%
\providecommand \translation [1]{[#1]}%
\providecommand \BibitemOpen [0]{}%
\providecommand \bibitemStop [0]{}%
\providecommand \bibitemNoStop [0]{.\EOS\space}%
\providecommand \EOS [0]{\spacefactor3000\relax}%
\providecommand \BibitemShut  [1]{\csname bibitem#1\endcsname}%
\let\auto@bib@innerbib\@empty
\bibitem [{\citenamefont {Loureiro}\ \emph {et~al.}(2007)\citenamefont
  {Loureiro}, \citenamefont {Schekochihin},\ and\ \citenamefont
  {Cowley}}]{Loureiro_Schekochihin_Cowley_07}%
  \BibitemOpen
  \bibfield  {author} {\bibinfo {author} {\bibfnamefont {N.~F.}\ \bibnamefont
  {Loureiro}}, \bibinfo {author} {\bibfnamefont {A.~A.}\ \bibnamefont
  {Schekochihin}}, \ and\ \bibinfo {author} {\bibfnamefont {S.~C.}\
  \bibnamefont {Cowley}},\ }\href {\doibase 10.1063/1.2783986} {\bibfield
  {journal} {\bibinfo  {journal} {Phys.Plasmas}\ }\textbf {\bibinfo {volume}
  {14}},\ \bibinfo {eid} {100703} (\bibinfo {year} {2007})}\BibitemShut
  {NoStop}%
\bibitem [{\citenamefont {Bhattacharjee}\ \emph {et~al.}(2009)\citenamefont
  {Bhattacharjee}, \citenamefont {Huang}, \citenamefont {Yang},\ and\
  \citenamefont {Rogers}}]{Bhattacharjee_Huang_Yang_Rogers_09}%
  \BibitemOpen
  \bibfield  {author} {\bibinfo {author} {\bibfnamefont {A.}~\bibnamefont
  {Bhattacharjee}}, \bibinfo {author} {\bibfnamefont {Y.-M.}\ \bibnamefont
  {Huang}}, \bibinfo {author} {\bibfnamefont {H.}~\bibnamefont {Yang}}, \ and\
  \bibinfo {author} {\bibfnamefont {B.}~\bibnamefont {Rogers}},\ }\href
  {\doibase 10.1063/1.3264103} {\bibfield  {journal} {\bibinfo  {journal}
  {Phys. Plasmas}\ }\textbf {\bibinfo {volume} {16}},\ \bibinfo {pages}
  {112102} (\bibinfo {year} {2009})}\BibitemShut {NoStop}%
\bibitem [{\citenamefont {Huang}\ and\ \citenamefont
  {Bhattacharjee}(2010)}]{Huang_Bhattacharjee_10}%
  \BibitemOpen
  \bibfield  {author} {\bibinfo {author} {\bibfnamefont {Y.-M.}\ \bibnamefont
  {Huang}}\ and\ \bibinfo {author} {\bibfnamefont {A.}~\bibnamefont
  {Bhattacharjee}},\ }\href {\doibase 10.1063/1.3420208} {\bibfield  {journal}
  {\bibinfo  {journal} {Phys. Plasmas}\ }\textbf {\bibinfo {volume} {17}},\
  \bibinfo {pages} {062104} (\bibinfo {year} {2010})}\BibitemShut {NoStop}%
\bibitem [{\citenamefont {Biskamp}(1986)}]{Biskamp_86}%
  \BibitemOpen
  \bibfield  {author} {\bibinfo {author} {\bibfnamefont {D.}~\bibnamefont
  {Biskamp}},\ }\href {\doibase 10.1063/1.865670} {\bibfield  {journal}
  {\bibinfo  {journal} {Physics of Fluids}\ }\textbf {\bibinfo {volume} {29}},\
  \bibinfo {pages} {1520} (\bibinfo {year} {1986})}\BibitemShut {NoStop}%
\bibitem [{\citenamefont {Parker}\ \emph {et~al.}(1990)\citenamefont {Parker},
  \citenamefont {Dewar},\ and\ \citenamefont
  {Johnson}}]{Parker_Dewar_Johnson_90}%
  \BibitemOpen
  \bibfield  {author} {\bibinfo {author} {\bibfnamefont {R.~D.}\ \bibnamefont
  {Parker}}, \bibinfo {author} {\bibfnamefont {R.~L.}\ \bibnamefont {Dewar}}, \
  and\ \bibinfo {author} {\bibfnamefont {J.~L.}\ \bibnamefont {Johnson}},\
  }\href {\doibase 10.1063/1.859340} {\bibfield  {journal} {\bibinfo  {journal}
  {Phys. Fluids B}\ }\textbf {\bibinfo {volume} {2}},\ \bibinfo {pages} {508}
  (\bibinfo {year} {1990})}\BibitemShut {NoStop}%
\bibitem [{\citenamefont {Lapenta}(2008)}]{Lapenta_08}%
  \BibitemOpen
  \bibfield  {author} {\bibinfo {author} {\bibfnamefont {G.}~\bibnamefont
  {Lapenta}},\ }\href {\doibase 10.1103/PhysRevLett.100.235001} {\bibfield
  {journal} {\bibinfo  {journal} {Phys. Rev. Lett.}\ }\textbf {\bibinfo
  {volume} {100}},\ \bibinfo {pages} {235001} (\bibinfo {year}
  {2008})}\BibitemShut {NoStop}%
\bibitem [{\citenamefont {Daughton}\ \emph {et~al.}(2009)\citenamefont
  {Daughton}, \citenamefont {Roytershteyn}, \citenamefont {Albright},
  \citenamefont {Karimabadi}, \citenamefont {Yin},\ and\ \citenamefont
  {Bowers}}]{Daughton_etal_09}%
  \BibitemOpen
  \bibfield  {author} {\bibinfo {author} {\bibfnamefont {W.}~\bibnamefont
  {Daughton}}, \bibinfo {author} {\bibfnamefont {V.}~\bibnamefont
  {Roytershteyn}}, \bibinfo {author} {\bibfnamefont {B.~J.}\ \bibnamefont
  {Albright}}, \bibinfo {author} {\bibfnamefont {H.}~\bibnamefont
  {Karimabadi}}, \bibinfo {author} {\bibfnamefont {L.}~\bibnamefont {Yin}}, \
  and\ \bibinfo {author} {\bibfnamefont {K.~J.}\ \bibnamefont {Bowers}},\
  }\href {\doibase 10.1103/PhysRevLett.103.065004} {\bibfield  {journal}
  {\bibinfo  {journal} {Phys. Rev. Lett.}\ }\textbf {\bibinfo {volume} {103}},\
  \bibinfo {pages} {065004} (\bibinfo {year} {2009})}\BibitemShut {NoStop}%
\bibitem [{\citenamefont {Cassak}\ \emph {et~al.}(2009)\citenamefont {Cassak},
  \citenamefont {Shay},\ and\ \citenamefont {Drake}}]{Cassak_Shay_Drake_09}%
  \BibitemOpen
  \bibfield  {author} {\bibinfo {author} {\bibfnamefont {P.~A.}\ \bibnamefont
  {Cassak}}, \bibinfo {author} {\bibfnamefont {M.~A.}\ \bibnamefont {Shay}}, \
  and\ \bibinfo {author} {\bibfnamefont {J.~F.}\ \bibnamefont {Drake}},\ }\href
  {\doibase 10.1063/1.3274462} {\bibfield  {journal} {\bibinfo  {journal}
  {Phys. Plasmas}\ }\textbf {\bibinfo {volume} {16}},\ \bibinfo {pages}
  {120702} (\bibinfo {year} {2009})}\BibitemShut {NoStop}%
\bibitem [{\citenamefont {{Loureiro}}\ \emph {et~al.}(2009)\citenamefont
  {{Loureiro}}, \citenamefont {{Uzdensky}}, \citenamefont {{Schekochihin}},
  \citenamefont {{Cowley}},\ and\ \citenamefont {{Yousef}}}]{Loureiro_etal_09}%
  \BibitemOpen
  \bibfield  {author} {\bibinfo {author} {\bibfnamefont {N.~F.}\ \bibnamefont
  {{Loureiro}}}, \bibinfo {author} {\bibfnamefont {D.~A.}\ \bibnamefont
  {{Uzdensky}}}, \bibinfo {author} {\bibfnamefont {A.~A.}\ \bibnamefont
  {{Schekochihin}}}, \bibinfo {author} {\bibfnamefont {S.~C.}\ \bibnamefont
  {{Cowley}}}, \ and\ \bibinfo {author} {\bibfnamefont {T.~A.}\ \bibnamefont
  {{Yousef}}},\ }\href {\doibase 10.1111/j.1745-3933.2009.00742.x} {\bibfield
  {journal} {\bibinfo  {journal} {Mon. Not. R. Astron. Soc.}\ }\textbf
  {\bibinfo {volume} {399}},\ \bibinfo {pages} {L146} (\bibinfo {year}
  {2009})}\BibitemShut {NoStop}%
\bibitem [{\citenamefont {Hahm}\ and\ \citenamefont
  {Kulsrud}(1985)}]{Hahm_Kulsrud_85}%
  \BibitemOpen
  \bibfield  {author} {\bibinfo {author} {\bibfnamefont {T.~S.}\ \bibnamefont
  {Hahm}}\ and\ \bibinfo {author} {\bibfnamefont {R.~M.}\ \bibnamefont
  {Kulsrud}},\ }\href {\doibase 10.1063/1.865247} {\bibfield  {journal}
  {\bibinfo  {journal} {Phys. Fluids}\ }\textbf {\bibinfo {volume} {28}},\
  \bibinfo {pages} {2412} (\bibinfo {year} {1985})}\BibitemShut {NoStop}%
\bibitem [{\citenamefont {Uzdensky}\ \emph {et~al.}(1996)\citenamefont
  {Uzdensky}, \citenamefont {Kulsrud},\ and\ \citenamefont
  {Yamada}}]{Uzdensky_Kulsrud_Yamada_96}%
  \BibitemOpen
  \bibfield  {author} {\bibinfo {author} {\bibfnamefont {D.~A.}\ \bibnamefont
  {Uzdensky}}, \bibinfo {author} {\bibfnamefont {R.~M.}\ \bibnamefont
  {Kulsrud}}, \ and\ \bibinfo {author} {\bibfnamefont {M.}~\bibnamefont
  {Yamada}},\ }\href {\doibase 10.1063/1.871746} {\bibfield  {journal}
  {\bibinfo  {journal} {Phys. Plasmas}\ }\textbf {\bibinfo {volume} {3}},\
  \bibinfo {pages} {1220} (\bibinfo {year} {1996})}\BibitemShut {NoStop}%
\bibitem [{\citenamefont {Uzdensky}\ and\ \citenamefont
  {Kulsrud}(1997)}]{Uzdensky_Kulsrud_97}%
  \BibitemOpen
  \bibfield  {author} {\bibinfo {author} {\bibfnamefont {D.~A.}\ \bibnamefont
  {Uzdensky}}\ and\ \bibinfo {author} {\bibfnamefont {R.~M.}\ \bibnamefont
  {Kulsrud}},\ }\href {\doibase 10.1063/1.872538} {\bibfield  {journal}
  {\bibinfo  {journal} {Phys. Plasmas}\ }\textbf {\bibinfo {volume} {4}},\
  \bibinfo {pages} {3960} (\bibinfo {year} {1997})}\BibitemShut {NoStop}%
\bibitem [{\citenamefont {Uzdensky}\ and\ \citenamefont
  {Kulsrud}(2000)}]{Uzdensky_Kulsrud_00}%
  \BibitemOpen
  \bibfield  {author} {\bibinfo {author} {\bibfnamefont {D.~A.}\ \bibnamefont
  {Uzdensky}}\ and\ \bibinfo {author} {\bibfnamefont {R.~M.}\ \bibnamefont
  {Kulsrud}},\ }\href {\doibase 10.1063/1.1308081} {\bibfield  {journal}
  {\bibinfo  {journal} {Phys. Plasmas}\ }\textbf {\bibinfo {volume} {7}},\
  \bibinfo {pages} {4018} (\bibinfo {year} {2000})}\BibitemShut {NoStop}%
\bibitem [{\citenamefont {Yamada}\ \emph {et~al.}(2010)\citenamefont {Yamada},
  \citenamefont {Kulsrud},\ and\ \citenamefont {Ji}}]{Yamada_Kulsrud_Ji_10}%
  \BibitemOpen
  \bibfield  {author} {\bibinfo {author} {\bibfnamefont {M.}~\bibnamefont
  {Yamada}}, \bibinfo {author} {\bibfnamefont {R.}~\bibnamefont {Kulsrud}}, \
  and\ \bibinfo {author} {\bibfnamefont {H.}~\bibnamefont {Ji}},\ }\href
  {\doibase 10.1103/RevModPhys.82.603} {\bibfield  {journal} {\bibinfo
  {journal} {Rev. Mod. Phys.}\ }\textbf {\bibinfo {volume} {82}},\ \bibinfo
  {pages} {603} (\bibinfo {year} {2010})}\BibitemShut {NoStop}%
\bibitem [{\citenamefont {Wang}\ and\ \citenamefont
  {Bhattacharjee}(1995)}]{Wang_Bhattacharjee_95}%
  \BibitemOpen
  \bibfield  {author} {\bibinfo {author} {\bibfnamefont {X.}~\bibnamefont
  {Wang}}\ and\ \bibinfo {author} {\bibfnamefont {A.}~\bibnamefont
  {Bhattacharjee}},\ }\href {\doibase 10.1063/1.871088} {\bibfield  {journal}
  {\bibinfo  {journal} {Phys. Plasmas}\ }\textbf {\bibinfo {volume} {2}},\
  \bibinfo {pages} {171} (\bibinfo {year} {1995})}\BibitemShut {NoStop}%
\bibitem [{\citenamefont {Boozer}\ and\ \citenamefont
  {Pomphrey}(2010)}]{Boozer_Pomphrey_11}%
  \BibitemOpen
  \bibfield  {author} {\bibinfo {author} {\bibfnamefont {A.~H.}\ \bibnamefont
  {Boozer}}\ and\ \bibinfo {author} {\bibfnamefont {N.}~\bibnamefont
  {Pomphrey}},\ }\href {\doibase 10.1063/1.3507307} {\bibfield  {journal}
  {\bibinfo  {journal} {Phys. Plasmas}\ }\textbf {\bibinfo {volume} {17}},\
  \bibinfo {pages} {110707} (\bibinfo {year} {2010})}\BibitemShut {NoStop}%
\bibitem [{\citenamefont {McGann}\ \emph {et~al.}(2010)\citenamefont {McGann},
  \citenamefont {Hudson}, \citenamefont {Dewar},\ and\ \citenamefont {{von
  Nessi}}}]{McGann_Hudson_Dewar_vonNessi_10}%
  \BibitemOpen
  \bibfield  {author} {\bibinfo {author} {\bibfnamefont {M.}~\bibnamefont
  {McGann}}, \bibinfo {author} {\bibfnamefont {S.~R.}\ \bibnamefont {Hudson}},
  \bibinfo {author} {\bibfnamefont {R.~L.}\ \bibnamefont {Dewar}}, \ and\
  \bibinfo {author} {\bibfnamefont {G.}~\bibnamefont {{von Nessi}}},\ }\href
  {\doibase 10.1016/j.physleta.2010.06.014} {\bibfield  {journal} {\bibinfo
  {journal} {Phys. Letts. A}\ }\textbf {\bibinfo {volume} {374}},\ \bibinfo
  {pages} {3308} (\bibinfo {year} {2010})}\BibitemShut {NoStop}%
\bibitem [{Note1()}]{Note1}%
  \BibitemOpen
  \bibinfo {note} {Hahm and Kulsrud also considered the general solution
  $\protect \mathaccentV {tilde}07E\psi = A\psi _{\protect \rm II}+ (1-A)\psi
  _{\protect \rm I}$, taking the constant $A(t)$ [expressed in terms of the
  reconnected flux $\alpha \psi _{\protect \rm II}(0,0) A(t)$] to represent the
  time evolution of the reconnection process.}\BibitemShut {Stop}%
\bibitem [{\citenamefont {Milne-Thomson}(1968)}]{Milne-Thomson_68}%
  \BibitemOpen
  \bibfield  {author} {\bibinfo {author} {\bibfnamefont {L.~M.}\ \bibnamefont
  {Milne-Thomson}},\ }\href {http://books.google.com/books?id=cXcfyei9H4MC}
  {\emph {\bibinfo {title} {Theoretical Hydrodynamics}}},\ \bibinfo {edition}
  {5th}\ ed.,\ Dover Books on Physics Series\ (\bibinfo  {publisher} {Dover
  paperback/MacMillan hardback},\ \bibinfo {address} {London},\ \bibinfo {year}
  {1968})\ \bibinfo {note} {see Chapters V to XV for the use of complex
  variable methods in 2-dimensional hydrodynamics.}\BibitemShut {Stop}%
\bibitem [{\citenamefont {Syrovatsky}(1971)}]{Syrovatsky_71}%
  \BibitemOpen
  \bibfield  {author} {\bibinfo {author} {\bibfnamefont {S.~I.}\ \bibnamefont
  {Syrovatsky}},\ }\href@noop {} {\bibfield  {journal} {\bibinfo  {journal}
  {Sov. Phys. JETP}\ }\textbf {\bibinfo {volume} {33}},\ \bibinfo {pages} {933}
  (\bibinfo {year} {1971})}\BibitemShut {NoStop}%
\bibitem [{\citenamefont {Kulsrud}(2011)}]{Kulsrud_11}%
  \BibitemOpen
  \bibfield  {author} {\bibinfo {author} {\bibfnamefont {R.~M.}\ \bibnamefont
  {Kulsrud}},\ }\href {\doibase 10.1063/1.3628312} {\bibfield  {journal}
  {\bibinfo  {journal} {Phys. Plasmas}\ }\textbf {\bibinfo {volume} {18}},\
  \bibinfo {pages} {111201} (\bibinfo {year} {2011})}\BibitemShut {NoStop}%
\bibitem [{\citenamefont {Taylor}(1986)}]{Taylor_86}%
  \BibitemOpen
  \bibfield  {author} {\bibinfo {author} {\bibfnamefont {J.~B.}\ \bibnamefont
  {Taylor}},\ }\href {\doibase 10.1103/RevModPhys.58.741} {\bibfield  {journal}
  {\bibinfo  {journal} {Rev. Mod. Phys.}\ }\textbf {\bibinfo {volume} {58}},\
  \bibinfo {pages} {741} (\bibinfo {year} {1986})}\BibitemShut {NoStop}%
\bibitem [{\citenamefont {Bhattacharjee}\ and\ \citenamefont
  {Dewar}(1982)}]{Bhattacharjee_Dewar_82}%
  \BibitemOpen
  \bibfield  {author} {\bibinfo {author} {\bibfnamefont {A.}~\bibnamefont
  {Bhattacharjee}}\ and\ \bibinfo {author} {\bibfnamefont {R.~L.}\ \bibnamefont
  {Dewar}},\ }\href {\doibase 10.1063/1.863819} {\bibfield  {journal} {\bibinfo
   {journal} {Phys. Fluids}\ }\textbf {\bibinfo {volume} {25}},\ \bibinfo
  {pages} {887} (\bibinfo {year} {1982})}\BibitemShut {NoStop}%
\bibitem [{\citenamefont {Hudson}\ \emph {et~al.}(2012)\citenamefont {Hudson},
  \citenamefont {Dewar}, \citenamefont {Dennis}, \citenamefont {Hole},
  \citenamefont {McGann}, \citenamefont {von Nessi},\ and\ \citenamefont
  {Lazerson}}]{Hudson_etal_12b}%
  \BibitemOpen
  \bibfield  {author} {\bibinfo {author} {\bibfnamefont {S.~R.}\ \bibnamefont
  {Hudson}}, \bibinfo {author} {\bibfnamefont {R.~L.}\ \bibnamefont {Dewar}},
  \bibinfo {author} {\bibfnamefont {G.}~\bibnamefont {Dennis}}, \bibinfo
  {author} {\bibfnamefont {M.~J.}\ \bibnamefont {Hole}}, \bibinfo {author}
  {\bibfnamefont {M.}~\bibnamefont {McGann}}, \bibinfo {author} {\bibfnamefont
  {G.}~\bibnamefont {von Nessi}}, \ and\ \bibinfo {author} {\bibfnamefont
  {S.}~\bibnamefont {Lazerson}},\ }\href {\doibase 10.1063/1.4765691}
  {\bibfield  {journal} {\bibinfo  {journal} {Phys. Plasmas}\ }\textbf
  {\bibinfo {volume} {19}},\ \bibinfo {pages} {112502} (\bibinfo {year}
  {2012})}\BibitemShut {NoStop}%
\bibitem [{\citenamefont {Yoshida}\ and\ \citenamefont
  {Dewar}(2012)}]{Yoshida_Dewar_12}%
  \BibitemOpen
  \bibfield  {author} {\bibinfo {author} {\bibfnamefont {Z.}~\bibnamefont
  {Yoshida}}\ and\ \bibinfo {author} {\bibfnamefont {R.~L.}\ \bibnamefont
  {Dewar}},\ }\href {\doibase 10.1088/1751-8113/45/36/365502} {\bibfield
  {journal} {\bibinfo  {journal} {J. Phys. A: Math. Gen.}\ }\textbf {\bibinfo
  {volume} {45}},\ \bibinfo {pages} {365502} (\bibinfo {year}
  {2012})}\BibitemShut {NoStop}%
\bibitem [{\citenamefont {Kruskal}\ and\ \citenamefont
  {Kulsrud}(1958)}]{Kruskal_Kulsrud_58}%
  \BibitemOpen
  \bibfield  {author} {\bibinfo {author} {\bibfnamefont {M.~D.}\ \bibnamefont
  {Kruskal}}\ and\ \bibinfo {author} {\bibfnamefont {R.~M.}\ \bibnamefont
  {Kulsrud}},\ }\href {\doibase 10.1063/1.1705884} {\bibfield  {journal}
  {\bibinfo  {journal} {Phys. Fluids}\ }\textbf {\bibinfo {volume} {1}},\
  \bibinfo {pages} {265} (\bibinfo {year} {1958})}\BibitemShut {NoStop}%
\bibitem [{\citenamefont {{Wolfram Research, Inc.}}(2013)}]{Mathematica9}%
  \BibitemOpen
  \bibfield  {author} {\bibinfo {author} {\bibnamefont {{Wolfram Research,
  Inc.}}},\ }\href {http://support.wolfram.com/kb/472} {\emph {\bibinfo {title}
  {Mathematica, Version 9.0}}}\ (\bibinfo  {publisher} {Wolfram Research},\
  \bibinfo {address} {Champaign, Illinois, USA},\ \bibinfo {year}
  {2013})\BibitemShut {NoStop}%
\end{thebibliography}
%

\end{document}